\documentstyle[psfig]{mn}
\psfull

\def\ltsima{$\; \buildrel < \over \sim \;$}
\def\lsim{\lower.5ex\hbox{\ltsima}}
\def\gtsima{$\; \buildrel > \over \sim \;$}
\def\gsim{\lower.5ex\hbox{\gtsima}}
\def\mes{M\'esz\'aros}
\def\aver#1{\langle #1 \rangle}

\begin{document}

\title[Constraints on $\Gamma$ in GRB internal shocks]
{Constraints on the bulk Lorentz factor in the internal 
shock scenario for gamma--ray bursts}

\author[Lazzati, Ghisellini \& Celotti]
{Davide Lazzati$^{1,2}$, Gabriele Ghisellini$^1$ \& Annalisa Celotti$^3$\\
$^1$ Osservatorio Astronomico di Brera, Via Bianchi 46, I--23807
Merate (Lc), Italy \\
$^2$ Dipartimento di Fisica, Universit\`a degli Studi di Milano,
Via Celoria 16, I--20133 Milano, Italy \\
$^3$ S.I.S.S.A., Via Beirut 2/4, I--34014 Trieste, Italy\\
E--mail: {\tt lazzati@merate.mi.astro.it, gabriele@merate.mi.astro.it, 
celotti@sissa.it}}
\maketitle

\begin{abstract}
We investigate, independently of specific emission models, the
constraints on the value of the bulk Lorentz factor $\Gamma$ of a
fireball. We assume that the burst emission comes from internal shocks
in a region transparent to Thomson scattering and before deceleration
due to the swept up external matter is effective.  
We consider the role of Compton drag in decelerating fast moving 
shells before they interact with slower ones, thus limiting the 
possible differences in bulk Lorentz factor of shells.  
Tighter constraints on the possible
range of $\Gamma$ are derived by requiring that the internal shocks
transform more than a few per cent of the bulk energy into radiation.
Efficient bursts may require a hierarchical scenario, where a shell
undergoes multiple interactions with other shells.  We conclude that
fireballs with average Lorentz factors larger than 1000 are unlikely
to give rise to the observed bursts.
\end{abstract}

\section{introduction}
\label{uno}

In the last few years, increasing evidence in favor of the fireball
model (Rees \& \mes~1992) for gamma--ray bursts has been gathered, thanks to 
the observations of {\it Beppo}SAX (Boella et al. 1997). 
The power law decay of the optical afterglow of several bursts has lasted 
for time--scales of a year, in striking agreement with the simplest 
fireball scenario (Wijers, Rees \& \mes~1997), in which a shock wave 
propagates in the interstellar medium (ISM), accelerating particles which 
then emit by the synchrotron process (Sari, Piran \& Narayan 1998).
However a single shock wave cannot account for both the
temporal behavior of the $\gamma$--ray emission (Fenimore et
al. 1999a) and for the requirement of a high efficiency in converting
its kinetic energy into radiation (Sari \& Piran 1997).  
In fact the time variability structure of
bursts appears to be constant during the whole of the high energy
emission (Fenimore et al. 1999b), while the deceleration of a shock slowed down by
interactions with the ISM would produce a time dilation between the
first and the last spike of the burst. Moreover the observed variability cannot 
be explained as the consequence of inhomogeneities in the ISM since a radiative
efficiency of less than $1\%$ would be expected (Sari \& Piran 1997;
see however Dermer, B\"ottcher \& Chiang 1999).

In the internal shock scenario, put forward by Rees \& \mes~(1994), the 
inner engine produces many relativistic expanding shells (or an unsteady wind) 
with a distribution of Lorentz factors centered on a mean value $\aver{\Gamma}$ and
width $\Delta\Gamma \sim \aver{\Gamma}$.
The burst radiation is produced through the dissipation occurring when a faster shell 
catches up a slower one. No matter the physical mechanism producing the observed 
photons, the inner (hidden) engine is responsible for the temporal 
structure of the observed burst through the time history
of the shell emission (Kobayashi, Piran \& Sari 1997). 
Even if we may have information on the time--scale of the shell ejection,
it is hard to estimate the bulk Lorentz factor $\aver{\Gamma}$ of the 
relativistic outflow.  This is however a critical
parameter to unveil the mechanism(s) that powers the outflow and
the radiating process that produces $\gamma$--ray photons.
In fact, the knowledge of $\aver{\Gamma}$ allows us to estimate
the amount of baryon loading of the fireball and the intrinsic
frequency of the emitted photons.  
The constraints related to the compactness problem (see e.g. Piran
1992) give a lower limit $\aver{\Gamma} \gsim 100$, while
an upper limit $\aver{\Gamma} \lsim 10^5$ is inferred from the
need of having the fireball opaque till the acceleration stage is
completed.
This poorly constrains the properties of the outflow, leaving a broad
interval for physically possible $\aver{\Gamma}$.

In this letter we analyze the kinematic evolution of an inhomogeneous 
fireball, in which the flow is approximated by discrete shells 
with different intrinsic properties (baryon load, energy and Lorentz factor)
and the outflow energy is dissipated through binary shell--shell collisions. 
We derive general constraints on the relativistic properties of the 
outflow both in the simplest internal shock scenario, that requires
a single interaction for each couple of shells, and in the case of 
hierarchically developed internal shocks, in which successive shell mergers
end up in a single (or a few) more massive shells, whose interaction with
the ISM produces the afterglow. The latter scenario
is strongly favored by the need of dissipating a significant fraction 
of ordered outflow energy (see also Kobayashi et al. 1997).


\section{Kinematic of the flow}
\label{due}

We describe the relativistic outflow as constituted of $N_{\rm S}$ 
shells,
each with its own relativistic Lorentz factor $\Gamma_{\rm i}$,
mass $M_{\rm i}$ and energy $E_{\rm i} = \Gamma_{\rm i} M_{\rm i} c^2$. 
The global properties of the flow are described by the averages 
$\aver{\Gamma}$, $\aver{M}$ and $\aver{E}$ and their 
dispersions, by the total duration of the ejection of shells
$T$ (as measured in the rest frame of the inner engine) and by the time interval 
between the ejection of two successive shells $\Delta t$.
If the ejection time of a typical shell equals
the time of quiescent phase between the production of two consecutive 
shells, $\Delta t= T/2N_{\rm S}$ (it is thus assumed that  
the time necessary to eject a fast or a slow shell
is the same, see however Panaitescu, Spada and \mes~1999).

The average mass, energy and relativistic factor are related as 
\begin{equation}
\aver{M} = {E \over N_{\rm S} \aver{\Gamma} c^2} \simeq 5.6 
\times 10^{-7} E_{52} \, \aver{\Gamma}_2^{-1} N_{\{\rm S,2\}}^{-1} \;\; M_\odot, 
\end{equation}
where $E = 10^{52} E_{52}$~erg is the total energy of the outflow
\footnote{Here and in the following we 
parameterize a quantity $Q$ as $Q=10^{\rm x} Q_{\rm x}$ and adopt CGS units.}.

\begin{figure*}
\psfig{figure=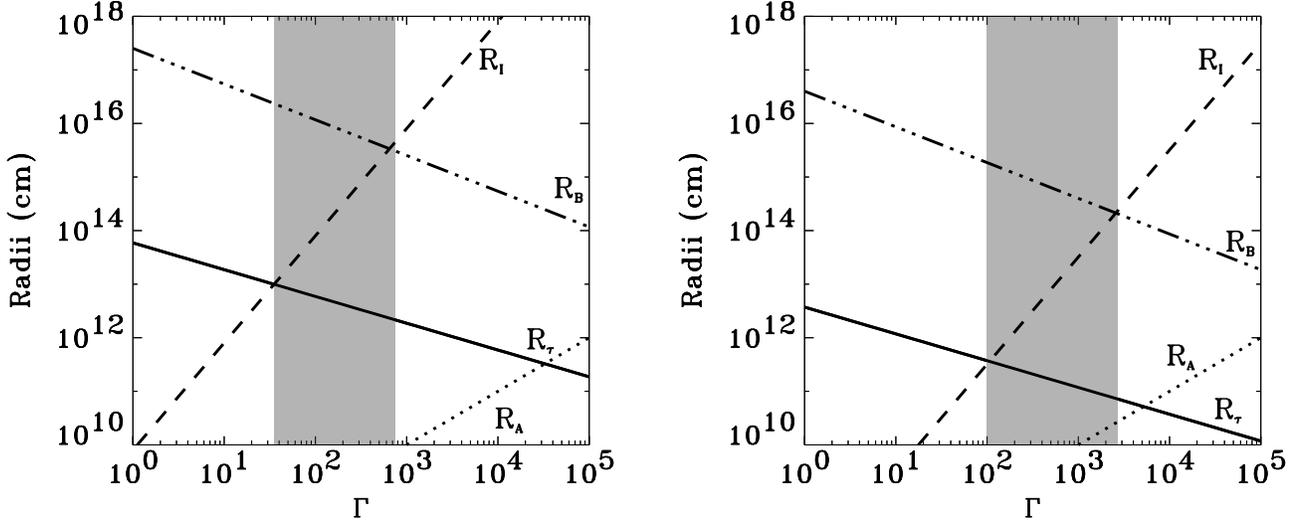,width=17.5cm}
\caption{{Value of the four {\rm relevant}  radii as a function of the
bulk average Lorentz factor $\aver{\Gamma}$ in a burst with energy $E=10^{52}$~erg
{\rm which lasts 10 seconds that produces} 100 shells (left panel) and
25000 shells (right panel). The ISM has been 
assumed uniform with density $n=1$~cm$^{-3}$. The shaded areas
correspond to the interval of $\aver{\Gamma}$ in which the four radii 
are correctly nested. 
If $\aver{\Gamma} < 35$ (100 in the right panel) the shells interact before 
becoming transparent and cannot emit the observed thin spectrum. 
If instead $\aver{\Gamma} > 750$ (2700 in the right panel) the shells are 
decelerated by the external medium before the mutual interaction 
occurs, producing an external and not an internal shock.}
\label{fi1}}
\end{figure*}

The dynamics of each shell is characterized by an initial phase in which 
it is accelerated to its final Lorentz factor as $\Gamma \sim R/R_0$, 
where $R_0$ is the distance of the shell from the center of expansion
at the ejection time (see e.g. Piran 1998). This phase ends when the final Lorentz
factor is reached at the `acceleration' radius:
\begin{equation}
R_{\rm A} = \Gamma_{\rm i} R_0 \simeq 10^9 \Gamma_{\{\rm i,2\}} R_{\{0,7\}}\,\, {\rm cm}. 
\end{equation}
During the following evolution, the shell coasts with constant velocity
until an interaction with another shell or with the ambient medium takes
place. Assuming that the acceleration phase lasts for a time interval 
negligible with respect to the coasting phase, the interaction between 
two shells with Lorentz factors $\Gamma_1$ and $\Gamma_2$ ($ > \Gamma_1$) 
and initially spaced by a time interval $\Delta t$ occurs at a radius:
\begin{equation}
R_I = 2 \, {\alpha_\Gamma^2 \over \alpha_\Gamma^2-1} \, \Gamma_1^2 c \Delta t = 
4.0 \times 10^{13} \Gamma_{\{1,2\}}^2 T_1 N_{\{\rm S,2\}}^{-1}\,\, {\rm cm}
\label{eq:ri}
\end{equation}
where the numerical value has been computed for 
$\alpha_\Gamma=\Gamma_2/\Gamma_1 =2$ 
and assuming $\Delta t=T/2 N_S$.

The interaction of the shells with the ambient medium is a continuous 
process and does not happen at a well determined radius. However
a typical scale can be estimated as the distance at which the  
shell Lorentz
factor is half its initial value, which is reached when the shell has swept up
an external mass $m$ equal to its rest mass divided by the initial 
Lorentz factor (see \mes~\& Rees 1997). Hence the deceleration radius can be 
approximated as ${M_i/\Gamma_i} = {(4/3)} \pi R_D^3 n m_p$, i.e.:
\begin{equation}
R_D \simeq 1.2 \times 10^{16} E_{52}^{1/3} n_0^{-1/3} \Gamma_{\{\rm i,2\}}^{-2/3} 
N_{\{\rm S,2\}}^{-1/3} 
\,\, {\rm cm}
\label{eq:rd}
\end{equation}
where $n$ is the density (assumed uniform) of the ambient medium and
shells of equal energy have been considered.  
Therefore, the more the shell is relativistic, the smaller the deceleration 
radius is (Eq.~\ref{eq:rd}), while, on the contrary,
the largest the radius at which internal shocks occur (see
Eq.~\ref{eq:ri}). This allows us to put some constraints
on the average relativistic factor of the flow, since observationally
the internal shocks set up before the afterglow, i.e. before the
development of the external shock.

To fully describe the kinematic of the flow, a fourth transition radius 
is important, i.e. the radius at which the shell becomes transparent (to Thomson 
scattering).
If, in fact, the shell becomes optically thin before the acceleration phase
is completed, the internal energy can escape from the shell and the 
acceleration process is damped. On the other hand, if two shells collide
when still opaque, they give rise to a single shell that is reaccelerated
to an intermediate Lorentz factor. The transparency radius corresponds to:
\begin{equation}
R_\tau = \left( {M_i \sigma_T \over 4 \pi m_p} \right)^{1/2}
\simeq 5.9 \times 10^{12} E_{52}^{1/2} \Gamma_{\{\rm i,2\}}^{-1/2} N_{\{\rm S,2\}}^{-1/2}
\,\, {\rm cm}
\label{eq:rt}
\end{equation}

\section{Comparing the radii}
\label{tre}

Fig.~\ref{fi1} shows a comparison of the four critical radii 
for a typical burst with energy $E=10^{52}$~erg and duration $T=$ 10 s.
The ISM has been assumed uniform with density $n=1$~cm$^{-3}$.
Two cases have been considered: the left panel refers to a burst made by a 
(relatively) small number of shells ($N_S=100$), while for the right 
panel $N_S=25000$. The smaller value has been derived assuming that 
each of the burst pulses fitted by Norris et al. (1996) is produced by the 
interaction of a shell pair. Norris et al. (1996) find a number of pulses
$N_p \lsim 50$, and hence we have $N_S = 2\, N_P \simeq 100$.
In the latter case, instead, the central engine emits shells at regular time 
intervals, equally spaced by the smallest variability time--scale 
observed in GRB light--curves ($200\mu$s, Schaefer \& Walker 1999).
In this situation each of the Norris et al. (1996) pulses
is considered as a blend of sub--pulses, with an envelope corresponding to the
varying efficiency of the central engine: 
indeed, Walker, Schaefer \& Fenimore (1999) have shown that
millisecond variability on top of the larger time--scale modulation is
a common feature of GRB light--curves.

In both cases the maximum value of $\aver{\Gamma}$ is bound by 
the requirement that internal shocks must happen before external ones, 
while the minimum value is constrained by the transparency condition.
This second constraint is less severe since, as already mentioned,  
collisions between opaque shells would simply cause a reacceleration 
of the merged shell. Note however that this could play a role in 
preventing the presence of very slow shells outside the transparency 
radius $R_\tau$. It should be also stressed that numerically the limit 
given by the transparency requirement is similar to the  
minimum value $\Gamma \gsim 100$, already
obtained from the compactness argument (e.g. Piran et al. 1996).
The acceleration radius does not impose any significant constraint, 
showing that from the kinematic and radiative points of view, shells could be
potentially accelerated even to very high Lorentz factors $\Gamma \sim 
10^5$.

From the condition $R_I < R_D$ we get an upper limit:
\begin{equation}
\Gamma_{\rm lim} = 800 \; T_1^{-3/8} \, E_{52}^{1/8} \, n_0^{-1/8} \, 
N_{\rm S,2}^{1/4}.
\label{gamlim}
\end{equation}
This limit is quite robust with respect to $E$ and $n$ 
while it is more dependent of $T$ and $N_S$. A variation of three
orders of magnitude of the total energy changes it by a factor $\sim 3$ only. 
The density of the ISM has been assumed uniform and $\sim$ one 
proton cm$^{-3}$. This is probably a lower limit. However, even in  the
case of the hypernova scenario (Paczynski 1998), in which GRB would occur in a 
much denser environment ($n \sim 10^4$~cm$^{-3}$), $\Gamma_{\rm lim}$ 
decreases only by a 
factor of $\sim 3$. Finally, the total burst duration of 10 seconds 
corresponds to the mean T90 parameter of the long GRBs, and is appropriate 
for the majority of bursts. 
Its small uncertainty influences $\Gamma_{\rm lim}$ rather weakly.

\section{Efficiency of internal shocks}
\label{qua}

\begin{figure}
\psfig{figure=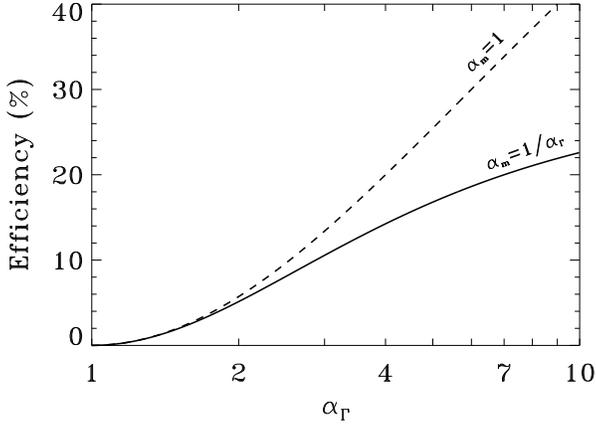,width=8cm}
\caption{{Radiative efficiency of internal shocks as a function
of the ratio between the relativistic $\Gamma$ factors of the 
interacting shells (Eq.~\ref{aneff}; $\Gamma_1, \Gamma_2 \gg 1$).
The solid curve refers to the situation in which the energies of the two
shells are comparable, i.e. $m_1 \, \Gamma_1 = m_2 \, \Gamma_2$, while 
the dashed curve shows the highest efficiency for a given 
$\alpha_\Gamma$, corresponding to shells of equal mass.
Since both shells have been assumed highly relativistic, the efficiency
does not depend on the value of $\Gamma_1$.}
\label{fi4}}
\end{figure}

The efficiency of internal shocks in converting the bulk outflow
energy into internal energy can be easily estimated since each collision 
satisfies energy and momentum conservation (see also Kobayashi et al. 1997). 
Consider two shells of rest masses $m_1$ and $m_2$ and  
Lorenz factors $\Gamma_1$ and $\Gamma_2$ ($> \Gamma_1$), respectively.
Calling $\epsilon$ the internal (random) energy of the merged shell
after the interaction, we have:
\begin{eqnarray} 
\Gamma_1 \, m_1 + \Gamma_2 \, m_2 &=& \Gamma_f \, \left(
m_1+m_2+\epsilon/c^2 \right) \nonumber \\
\Gamma_1 \, \beta_1 \, m_1 + \Gamma_2 \, \beta_2 \, m_2 &=&
\Gamma_f \, \beta_f \, \left(m_1+m_2+\epsilon/c^2 \right), 
\label{sist}
\end{eqnarray}
where the subscript $f$ refers to quantities after the interaction and 
$\Gamma = (1-\beta^2)^{-1/2}$.
If we assume that all of the internal energy is converted into
radiation, we obtain - independently of the emission mechanism - an
upper limit for the efficiency $\eta = \epsilon / (\Gamma_1 m_1 +
\Gamma_2 m_2)$. From the conservation equations 
an implicit solution for the final bulk Lorentz factor of the merged 
shells and the 
maximum radiative efficiency can then be derived:
\begin{eqnarray}
\beta_f &=& {\beta_1 + \alpha_\Gamma \, \alpha_m \, \beta_2 \over
1+\alpha_\Gamma \, \alpha_m} \nonumber \\
\eta &=& 1-{\Gamma_f \, (1+\alpha_m) \over 
\Gamma_1 \, (1+\alpha_\Gamma \, \alpha_m)},
\end{eqnarray}
where $\alpha_m = m_2/m_1$. 
The above relations give an {\it upper limit}
to the fraction of energy that can be radiated in photons. 
In fact, strictly speaking,
$\eta$ is the fraction of bulk kinetic energy converted into internal energy.
In the standard synchrotron shock model, this random energy is equally 
shared among
protons, electrons and magnetic field, and only one third of this energy
(the fraction going to electrons) can be radiated (see e.g. Panaitescu 
et al. 1999).
The major features of Eq.~\ref{sist} are that for a large difference in the 
Lorentz factors ($\alpha_\Gamma \gg 1$) the efficiency can approach unity 
and that for a fixed value of $\alpha_\Gamma$ the maximum efficiency is reached 
when $m_1=m_2$.
For $\Gamma_1, \Gamma_2 \gg 1$, as for internal shocks, the expression for
$\eta$ reduces to:
\begin{equation}
\eta \simeq 1 - {1 + \alpha_m \over \sqrt{ 
1+\alpha_m \left( \alpha_m + \alpha_\Gamma + 1/\alpha_\Gamma
\right)}}.
\label{aneff}
\end{equation}
An even simpler relation is found if the two shells have the same 
total energy (i.e. $\alpha_\Gamma = 1/\alpha_m$):
\begin{equation}
\eta \simeq 1 - {1+\alpha_m \over \sqrt{2+2\, \alpha_m^2}}.
\end{equation}
In this latter case the efficiency is always lower than 30\%, 
independently of $\alpha_\Gamma$. Fig.~\ref{fi4} shows the efficiency 
for low--intermediate values of $\alpha_\Gamma$ in two limits: the most 
efficient situation ($m_1=m_2$, dashed line) and the equal energy case 
($\Gamma_1 \, m_1 = \Gamma_2 \, m_2$, solid line). The limit $\Gamma_1,\Gamma_2\gg 1$ 
has been assumed. Note that, in this case, the efficiency does not depend on the 
value of $\Gamma_1$.

Let us now consider the global efficiency of the burst.
In the `standard' internal shock scenario (Rees \& \mes~1994) the shells 
have a distribution of Lorentz factors with mean value $\aver{\Gamma}$ and
width $\Delta\Gamma \sim \aver{\Gamma}$.
This in turn corresponds to a distribution of $\alpha_\Gamma$ with  
$\aver{\alpha_\Gamma} \sim 2$ within a factor of order unity 
(which accounts
for the particular shape of the distribution).
This implies (see Fig.~\ref{fi4}) $\eta \simeq 5\%$, independently of the 
mass ratio $\alpha_m$. A numerical simulation for a log--normal distribution
of $\Gamma$ with $\Delta \Gamma = \aver{\Gamma}$, gives $\eta = 6.8\%$.

These (upper limits on) efficiencies are problematic since, 
in the absence
of extreme beaming, the total energy release of the most
powerful GRBs would exceed by some orders of magnitude the maximum energy
release achievable by current models, that involve a stellar mass 
black hole as the energy source.
If, e.g., GRB~990123 had an efficiency of $5\%$, the required energy
would be $E_{\rm iso} \simeq 5 \times 10^{55}$~erg.
An extremely narrow beaming angle $\theta \simeq 0.5^\circ$ would be then 
needed to reduce the energy release to the value  $E_\theta = 10^{52}$~erg.
A higher efficiency can be only obtained by allowing for a broader distribution 
of relativistic factors $\Gamma$ and imposing that all shells have roughly the 
same mass, somehow requiring a finely tuned variable engine. Moreover, a very 
broad distribution of Lorentz factors implies that the Compton drag 
effect becomes crucial, as described in the following section.

\subsection{Compton drag}

Let us assume that the emission produced in the interaction between two shells
is isotropic in the shell comoving frame $K^\prime$.
Consider also a third shell, faster than the previous two, 
with Lorentz factor $\Gamma^\prime$ in that frame.
This  moves in the radiation bath of total energy
$E^\prime_{\rm ph}$ produced in the interaction, and thus
Compton scatters a fraction $\tau_T$ of the photons, 
increasing their energy by a factor $\Gamma^{\prime 2}$
(here $\tau_T$ is the Thomson optical depth of the fast shell).
The total energy lost by the fast shell in the frame $K^\prime$ is hence 
$\tau_T \, (\Gamma^{\prime 2}-1) E^\prime_{\rm ph}$.

If $\tau_T \, (\Gamma^{\prime 2}-1)$ approaches unity, there are two major 
consequences: first a significant part of the energy of photons 
is due to the inverse Compton mechanism rather than to the internal shock
dissipation process; second, the fast shell looses a significant
fraction of its bulk kinetic energy to up--scatter
the primary photons and this causes a braking of the fast shell (in the frame
$K^\prime$). Thus, even if the original distribution of $\Gamma$
factors were very broad, the early interactions would give rise to an
efficient inverse Compton drag on the faster shells, reducing the mean
value of $\alpha_\Gamma$ in the successive collisions and hence the
global efficiency of internal shocks.
Moreover, in this case the net energy produced through 
inverse Compton scattering
would be equal or even larger than that directly due to shocks, and the primary emission 
mechanism would be different. We plan to investigate this issue in a 
forthcoming paper.

\begin{figure}
\psfig{figure=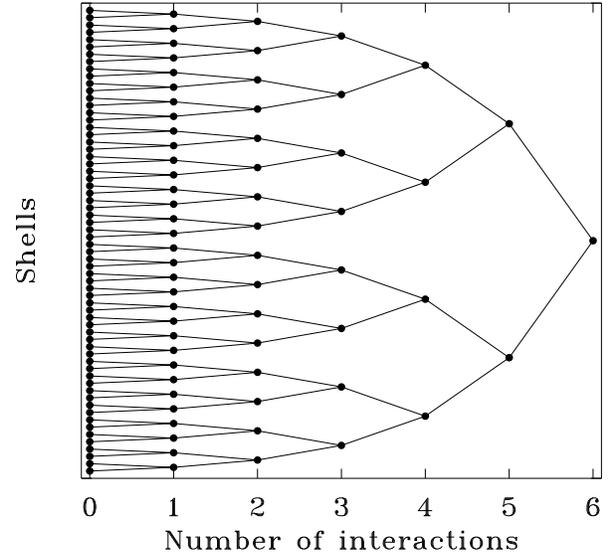,width=8cm}
\caption{{The simplest scheme for hierarchical internal shock.
The number of shells required to generate it 
with {\rm N} steps is given by $2^{\rm N}$.}
\label{treeps}}
\end{figure}

\section{Hierarchical internal shocks}
\label{cin}

The low efficiency predicted by Eq.~\ref{aneff} refers to a single interaction 
between two shells. However, each time this interaction takes place, a new shell 
is formed which can in turn catch up (or be caught by) another one. If the $i^{th}$ 
interaction has efficiency $\eta_i$, after $N$ of them the fraction of the 
initial bulk energy converted into photons will be:
\begin{equation}
\eta_{\rm N} = \sum_{i=1}^{N} \eta_i (1-\eta_{i-1}) ... (1-\eta_1)
\stackrel{\eta_i = \eta}{=\!\!=\!\!=\!\!\Longrightarrow}
\eta \sum_{i=1}^N (1-\eta)^{i-1}
\end{equation}
If the efficiency is constant ($\eta_i=\eta$) and $N$ is arbitrarily large, 
all the energy of the outflow can be converted into radiation, as 
$\eta_{_\infty} = 1$ for any value of $\eta$. 
For the typical efficiency $\eta = 6.8\%$ derived above, about 9 interactions
are  needed to obtain an overall $50\%$ efficiency.
Assuming that, as sketched in Fig.~\ref{treeps}, the collision `tree' develops 
as a binary bifurcation, the number $\mathcal{N}$ of shells that has to merge 
into a single one after $\sim$9 interactions is ${\mathcal N} = 2^9 = 512$.

This result has two main consequences. Firstly, it is unlikely that
a burst with high efficiency is produced by a small ($<$ 100) number of shells:
a scenario in which the inner engine emits a large number of shells is thus favored. Secondly, in deriving $\Gamma_{\rm lim}$ (Eq.~\ref{gamlim}) 
we assumed that a single interaction was enough to power a GRB. 
However, within the hierarchical scenario, a shell must interact with 
$\mathcal N$ other shells and hence Eq.~\ref{gamlim} must be modified as:
\begin{equation}
\Gamma_{\rm lim} = 800 \; T_1^{-3/8} \, E_{52}^{1/8} \, n_0^{-1/8} \, 
N_{\rm S,4}^{1/4} \, {\mathcal N}_2^{-1/4}
\label{gamlim2}
\end{equation}
which {gives $\Gamma_{\rm lim} \simeq 650$ for - say - $N_S=25000$ and ${\mathcal N}=512$}.
Even if this is an illustrative value only (the actual number depending 
on the details of the initial distribution of the Lorentz factors), 
we conclude that values of $\aver{\Gamma}$ larger than $\sim 1000$ 
are unlikely.

\section{Discussion}
\label{sei}

We have analyzed the kinematic efficiency of internal shocks
in a fireball made of many individual shells. Since the radius at which
the internal shocks set up grows with the average Lorentz factor of the 
flow, while the development of an external shock is favored by a higher
relativistic motion, we can put limits on the average Lorentz factor 
$\aver{\Gamma}$ of the flow if - as inferred from observations - internal shocks 
develop {\it before} the external one.
Moreover (see also Kobayashi et al. 1997) we conclude that a simple
internal shock scenario in which each shell is caught up
by a single faster shell, suffers from a very low radiative efficiency, unless
extremely different Lorentz factors are involved.
These would however cause a dramatic Compton drag effect, which in 
turn constrains the possibility of a very broad $\Gamma$ distribution.
A possible solution is to have a flow with moderately relativistic Lorentz 
factors $100 \lsim \aver{\Gamma} \lsim 600$ in which collisions develop 
until a single (or a few) massive shell is formed.
For this to happen the shells have to be ejected at small time intervals, 
of the order of milliseconds. In this case each burst is made of more than a 
thousand spikes that merge to produce the broad peaks often observed in GRB 
lightcurves. 
These peaks would then reflect a modulation of the hidden central 
engine that powers the burst, rather than a single collision.

It is interesting to ask whether an outflow with significantly higher Lorentz 
factors produces a burst or not. In a flow with very high $\aver{\Gamma}$, the 
internal shocks set-up at a larger radius, and hence a hierarchical internal shock
cannot develop. Since the efficiency is roughly proportional to the number of shocks 
a shell undergoes, a higher bulk gamma is linked to a lower efficiency.
Observationally, this implies that the ratio between burst and afterglow 
fluences is a function of $\aver{\Gamma}$, being larger for smaller $\aver{\Gamma}$.
A further consequence is that very short (millisecond) bursts should be 
characterized by smaller efficiencies and brighter afterglows.
To date, this can be only a prediction since the {\it Beppo}SAX trigger works 
only for long burst.

As a final note, following the explosion of the bright burst GRB~990123,
two values of $\aver{\Gamma}$ have been derived. 
Sari \& Piran (1999) obtain $\aver{\Gamma} \sim 200$ from the optical 
flash intensity while Liang et al. (1999) estimate  
$\aver{\Gamma} \sim 350$ from the lightcurve properties. 
Despite the differences, both values are in agreement with the general 
limits derived in this letter.

\section*{Acknowledgments}
We thanks S. Campana for stimulating comments and for carefully reading 
the manuscript.
DL and AC thank the Cariplo foundation and the Italian MURST for 
financial support, respectively. This research was supported in part 
by the National Science Foundation under Grant No. PHY94-07194 (AC).


\begin{thebibliography}{}
\bibitem{sdf} Boella G., Butler R. C., Perola G. C., Piro L., Scarsi L.,
	Bleeker J. A. M. 1997, A\&AS, 122, 299
\bibitem{fds} Dermer C.D., B\"ottcher M. \& Chiang J. 1999, ApJ, 512, L49
\bibitem{era} Fenimore E. E., Cooper C., Ramirez--Ruiz E., Sumner M. C.,
	Yoshida A., Namiki M. 1999a, ApJ, in press (astro-ph/9802200)
\bibitem{fen99} Fenimore E. E., Ramirez--Ruiz E.,
	Wu B. 1999b, ApJ, submitted (astro-ph/9902007)
\bibitem{ddd} Kobayashi S., Piran T. \& Sari R. 1997, ApJ, 490, 92
\bibitem{wsd} Liang E. P., Crider A., B\"ottcher M., Smith I.A. 1999,
	ApJ submitted (astro-ph/9903438)
\bibitem{sdg} \mes~P. \& Rees M. J. 1997, ApJ, 476, 232
\bibitem{nor96} Norris J. P. et al., 1996, ApJ, 459, 393
\bibitem{sgr} Paczynski B. 1998, ApJ, 494, L45
\bibitem{sva} Panaitescu A., Spada M. \& \mes~P. 1999, ApJ, 
submitted 
	(astro-ph/9905026)
\bibitem{pir98} Piran T. 1998, Physics Reports, (astro-ph/9810256)
\bibitem{res94}Rees M. J. \& \mes~P. 1994, ApJ, 430, L93
\bibitem{asd} Rees M. J. \& \mes~P. 1992, MNRAS, 258, 41
\bibitem{eer} Sari R. \& Piran T. 1997, ApJ, 485, 270
\bibitem{wet} Sari R., Piran T. \& Narayan R. 1998, ApJ, 497, L17
\bibitem{tsk} Sari R. \& Piran T. 1999, ApJ, submitted
(astro-ph/9902009)
\bibitem{sdj} Schaefer B. E. \& Walker K. C. 1999, ApJ, 511, L89
\bibitem{wer} Walker K. C., Schaefer B. E. \& Fenimore E. E. 1999,
	ApJ, submitted (astro-ph/9810271)
\bibitem{wij97} Wijers R. A. M. J., Rees M. J. \& \mes~P. 1997,
	MNRAS, 288, L51
\end{thebibliography}
\end{document}